\documentclass[twocolumn,aps,prl]{revtex4}

\usepackage{graphicx}
\usepackage{dcolumn}
\usepackage{bm}

\begin{document}

\title{Aging of the Nonlinear Optical Susceptibility of colloidal solutions}
\author{
Neda Ghofraniha$^1$, Claudio Conti$^{2,3}$ and Giancarlo
Ruocco$^{4,3}$ }

\affiliation{ $^{1}$SMC-INFM-CNR, c/o Universita' di Roma ``La
Sapienza'',
P. A. Moro 2, 00185, Roma, Italy\\
$^{2}$Research Center ``Enrico Fermi'', Via Panisperna 89/A,
00184 Rome, Italy \\
$^{3}$SOFT-INFM-CNR, c/o Universita' di Roma ``La Sapienza'',  P.
A. Moro 2, 00185, Roma, Italy\\
$^{4}$Diparrtimento di Fisica, Universita' di Roma ``La
Sapienza'', P. A. Moro 2, 00185, Roma, Italy }

\date{\today}

\begin{abstract}
Using Z-scan and dynamic light scattering measurements we
investigate the nonlinear optics response of a colloidal solution
undergoing dynamics slowing down with age. We study the high
optical nonlinearity of an organic dye (Rhodamine B) dispersed in
a water-clay (Laponite) solution, at different clay concentrations
(2.0 wt\% - 2.6 wt\%),  experiencing the gelation process. We
determine the clay platelets self diffusion coefficient and, by
its comparison with the structural relaxation time, we conclude
that  the gelation process proceeds through the structuring of
interconnecting clay platelets network rather than through
clusters growth and  aggregation.
\end{abstract}


 \maketitle

{\it Introduction.} Complex systems, including soft-materials, are
characterized by an exponentially large number of metastable
states \cite{MPVBook} routinely visited while the system is
evolving towards thermodynamic equilibrium if prepared far from
equilibrium. This process can extend on timescales much longer
than the atomic/molecular motions ones, due to the large number of
metastable states . As a result, the response features, like the
relaxation time, depend on the specific instant ({\it the waiting
time} $t_w$) and are determined by the current meta-stable state.
This mechanism is known as ``aging''. During aging, the
fluctuation-dissipation theory (FDT) does not strictly hold, but a
generalization has been recently proposed \cite{Cug93PRL} and
numerically tested \cite{Cug94JPA}. According to this
generalization the temperature, appearing in the FDT as
proportionality factor between Response and Correlation function,
is substituted by and effective temperature, higher that the
actual one. Consequently in aging systems, which are subject to
fluctuations larger than the corresponding equilibrium one, also
nonlinear effects are expected to be strongly sensible to the
presence of dynamics slowing down with age.

In this respect, nonlinear optics experiments in complex systems
can be considered as a promising tool to study these out of
equilibrium dynamics. However, if on one hand the temporal
behaviour of the nonlinear optical coefficients was considered
since the beginning of nonlinear optics \cite{ShenBook} and
various authors recently reported detailed investigations and new
models for the time dynamics of specific nonlinear optical effects
(like modulational instability or soliton formation,
\cite{Wright99,Conti05,Streppel05,DelRe06}). At present, only few
studies have been devoted to investigate the time evolution of
non-linear optics properties in soft matter. Specifically, in
Ref.~\cite{Rok05JPCM} the frequency dependence of $\chi_2$ has
been used to determine the influence of solvents on the molecular
arrangement of colloidal surface groups and how this affects the
effective interaction potentials controlling gelation. In
Ref.s~\cite{Pil02JPS,Yav04OM}, Z-scan measurements have been used
to follow the solvent evaporation in thin polymer films.

Very little is known about the microscopic process leading to the
arrested state; thus, with the aim of studying the gelation
process in colloidal materials, we use here the non-linear optical
response, as measured by the Z-scan technique, to determine the
self diffusion coefficient of the colloidal particles, a quantity
not directly accessible by other techniques. Specifically, in this
letter we report on an experimental investigation of a Laponite
clay powder dispersed in dye Rhodamine-B aqueous solution,
displaying both high thermal nonlinear optical response and
gelation process characteristic of the clay suspension. Clear
experimental evidence and quantitative assessment of aging of the
nonlinear susceptibility is reported.

{\it Sample preparation.} Rhodamine B (RhB) solution in deionized
water ({\it p}H=7) was prepared at 0.6 mM concentration. Laponite
powder, supplied by Laporte Ltd, was then dispersed in RhB
solution, stirred vigorously until the suspensions were cleared
and filtered through $0.22\;\mu$m pore size Millipore filters. We
consider the starting aging time ($t_w\,=\,0$) as the time when
the suspension was filtered. Four different samples at clay
concentrations between 2.0 wt\% and 2.6 wt\% were prepared and on
each of them the two different experiments (DLS and Z-scan)
described below were contemporarily performed.

{\it Evidences of aging.} In order to provide a clear evidence of
the non stationary dynamics in our samples, and for their
quantitative characterization, we resorted to dynamic light
scattering (DLS) measurements, which were performed using a
homemade correlator \cite{RdL} in combination with a standard
optical setup based on a He-Ne ($\lambda$=632.8 nm) 10 mw laser, a
monomode optical fiber and a photomultiplier detector. We observed
directly the normalized intensity $I(q,t)$ correlation function
$g_2(q,t)= \langle I(q,t)I(q,0) \rangle / \langle I(q,0)
\rangle^2$, with $q$ the modulus of the scattering wave vector
defined as $q=(4\,\pi\,n/\lambda)\,sin\,(\theta/2)$, being the
scattering angle $\theta=90^\circ$ in our experiment. DLS data
were fitted assuming a correlation function made by the sum of an
exponential function with relaxation time $\tau_1$ and a stretched
exponential function with relaxation time $\tau_2$ and stretching
coefficient $\beta$, as reported in \cite{Ruz04PRL} and commonly
used in photocorrelation measurements analysis for liquid-like
(ergodic) samples.

\begin{figure}[!h]
\begin{center}
\includegraphics[width=8.5cm] {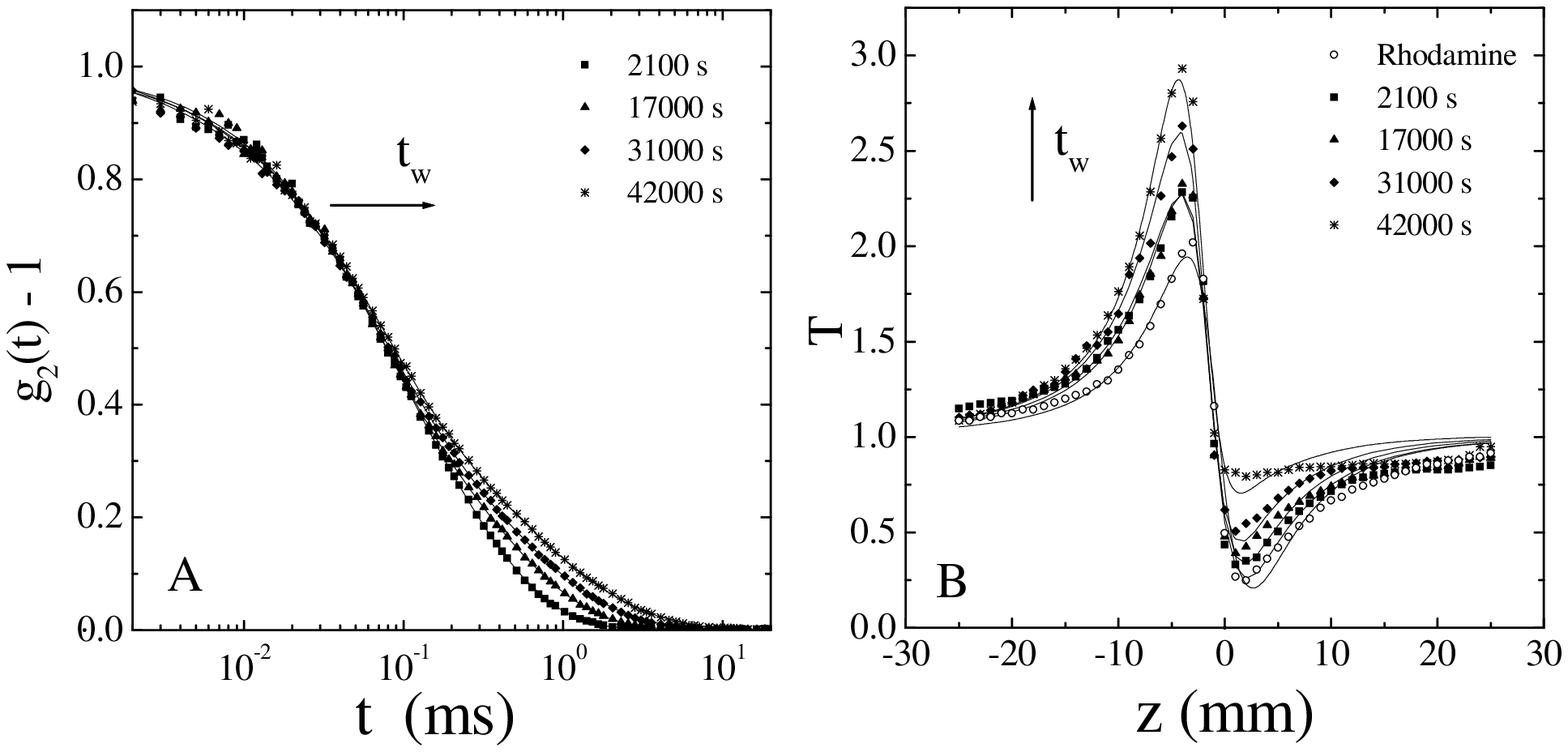}
\caption{(A) DLS homodyne correlation functions and (B) Z-scan
data of Laponite 2.2 wt \% dispersion in RhB  0.6 mM water
solution for different waiting times $t_w$ and corresponding fits
(full lines), see text. In (B) is also reported RhB 0.6 mM scan
(open circles). \label{fig1}}
\end{center}
\end{figure}

\begin{figure}[!h]
\begin{center}
\includegraphics[width=8.5cm] {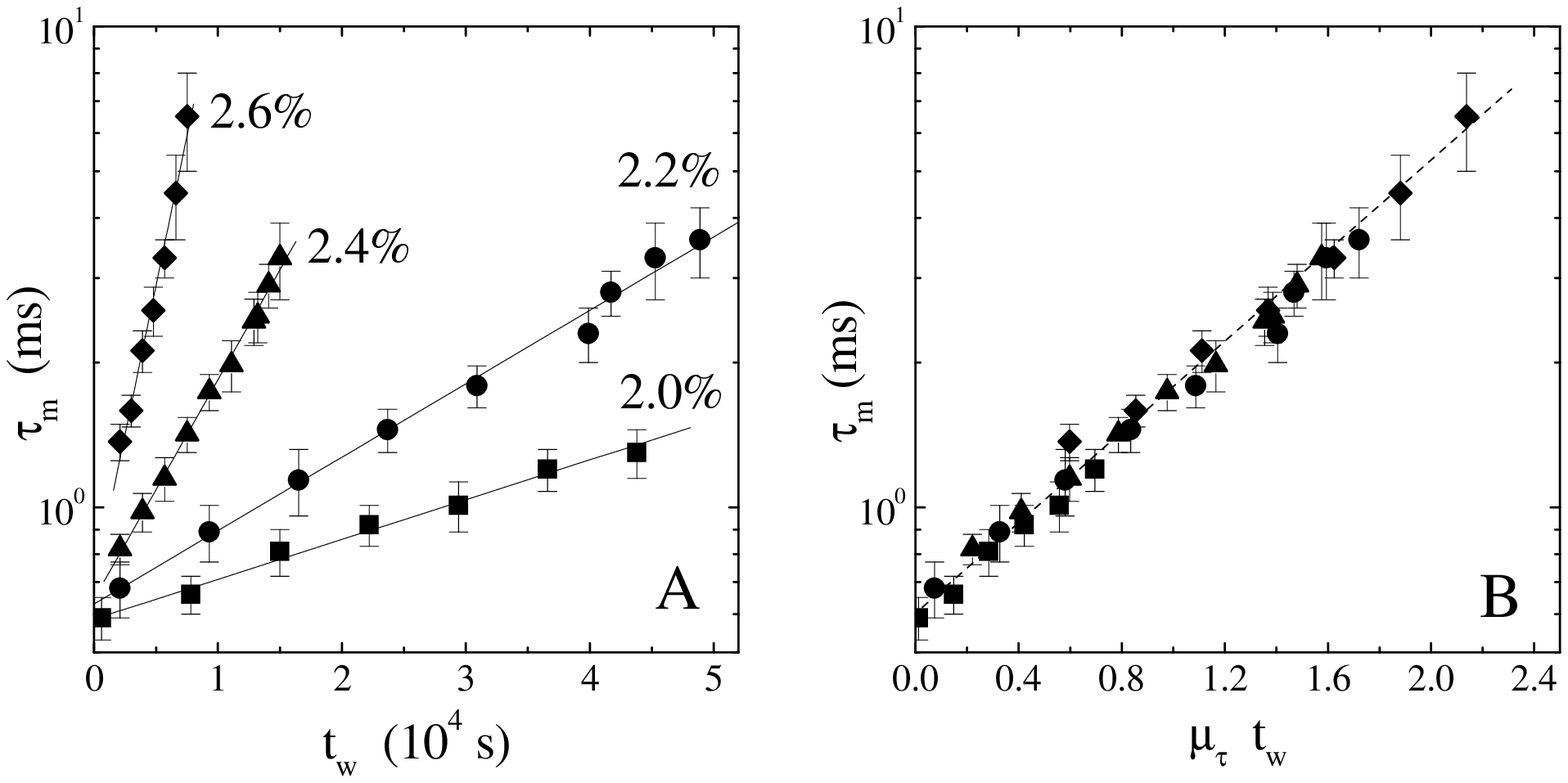}
\caption{Mean relaxation time $\tau_m$ vs. waiting time $t_w$ (A)
and vs. the scaled variable $\mu_\tau\, t_w$ (B) in log-linear
scale at various clay concentrations; continous lines in (A) are
the fitted curves. \label{fig2}}
\end{center}
\end{figure}

Example of DLS data, together with their best fits, are shown in
Fig.~\ref{fig1}A as symbols and full lines at the indicated
waiting times. The mean relaxation time $\tau_m$
\begin{equation}\label{taum}
\tau_m=\tau_2\,{\beta^{-1}}\,\Gamma\,\left({\beta^{-1}} \right),
\end{equation}
with $\Gamma(x)$ the usual Euler gamma function, can be taken as a
parameter representing the slow dynamics behaviour of the
correlation function, while the fast dynamics correlation time
$\tau_1$ turns out to be constant for our clay concentrations
during the gelation process, as described elsewhere
\cite{Ruz04JPHYSCM}. In Fig.~\ref{fig2}A  the $\tau_m$ vs. $t_w$
data in log-linear scale highlight the common exponential growth
behaviour $\tau_m(t_w) = \tau_0\,\mbox{exp}\,(\mu_\tau\, t_w)$
(full line trough the data) of the mean relaxation time for
different Laponite concentrations. These results reflect clearly
the aging behaviour of the material, that corresponds to $t_w$
dependent relaxation times. The specific $t_w$ dependence of
$\tau_m(t_w)$ leads to consider a scaling law, that makes all data
collapse on a single master curve. This curve is reported in
Fig.~\ref{fig2}B where the the DLS derived $\tau_m(t_w)$ data are
shown as a function of $\mu_\tau t_w$.

\begin{figure}[!h]
\begin{center}
\includegraphics[width=7cm] {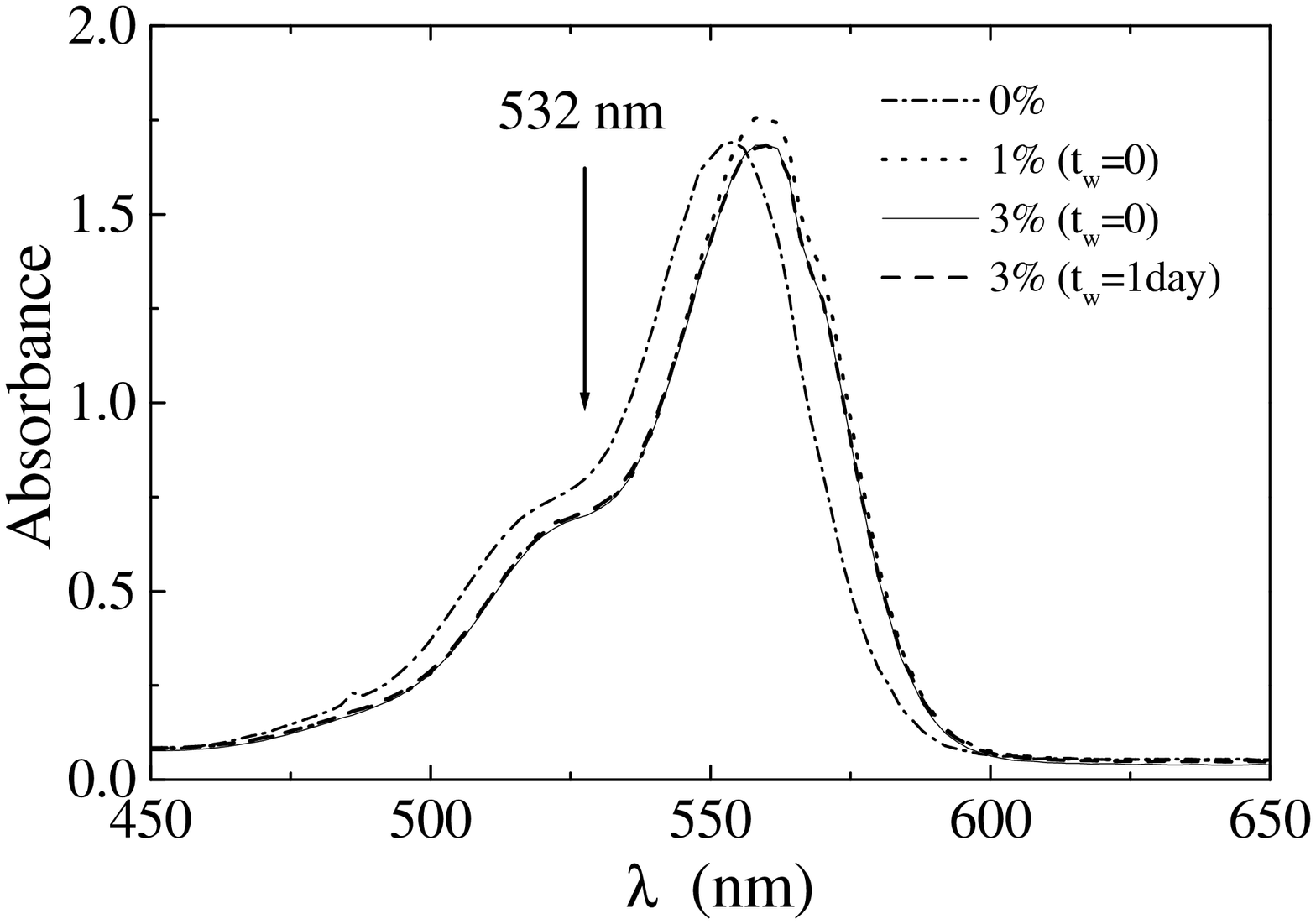}
\caption{Absorption spectra of RhB $2.4\,10^{-5}\; M$ water
solution ($0\,\%$) and Laponite dispersions in RhB $2.4\,10^{-5}\;
M$ solution at diffrent clay concentrations and waiting times.
\label{fig3}}
\end{center}
\end{figure}

{\it Linear absorption measurements.} In order to show that aging
has no role in the {\it linear} optical response of our system, we
measured the linear absorption spectra by UV-visible spectrometer.
In Fig.~\ref{fig3} we report representative absorption spectra of
RhB $2.4\times 10^{-5}$ M water solution at 0 wt \%, 1 wt \% and 3
wt \% clay concentrations, showing that the linear absorption of
RhB changes in presence of Laponite and it can be considered
independent on colloid concentration and on the waiting time.
These samples were prepared following the same procedure described
above, using a lower concentrated RhB solution to avoid saturation
in absorbance spectra. The linear absorbance coefficient
$\alpha_0$,  for the 0.6 mM Rhb solution was also found to be
$\alpha_0\simeq 6.2$ mm$^{-1}$ and $\alpha_0\simeq 4.8$ mm$^{-1}$
for all our RhB-Laponite dispersions, by measuring the sample
transmission at $\lambda=532$ nm confirming the absence of
aging in the linear optical absorption.

{\it Aging of the nonlinear susceptibility}. Using the Z-scan
technique \cite{Bah90a} we measured both nonlinear absorption and
refraction: the sample was moved along the $z-$axis through the
focal point of the input Gaussian beam, and the transmitted power
was measured as a function of $z$ in the far field using a
photodiode behind a small calibrated pinhole. In our experiments
we used a CW pumped diode laser operating at power $P$=10 mW and
wavelenght $\lambda$=532 nm. The beam was focused by means of a 75
mm focal length lens giving a 20 $\mu$m beam waist radius $w_o$
and a $I_0=8\times 10^6$ W/m$^2$ beam central intensity at the
focus ($z=0$). A photodetector, at distance 310 mm from lens
focus, was used to probe the light power behind a 2 mm aperture
and the sample was scanned across the focus with a 5 cm
translation stage.
In Fig.~\ref{fig1}B we show as an example the Z-scan curves of
RhB- 2.2 wt \% clay dispersion at increasing $t_w$ correspondingly
to the DLS correlation functions (shown in panel A). The dynamics
slowing down of the system clearly affects the nonlinear
susceptibility. In Fig.~\ref{fig1}B the 0.6 mM Rhb solution (no
clay) transmittance curve (open circles) is also reported. Notice
that, similarly to the linear absorption shown in Fig.~\ref{fig3},
the nonlinear optical response is influenced by the presence of
the Laponite platlets, while, more interestingly, contrarily to
the linear absorbance, the nonlinear susceptibility does actually
vary undergoing aging.

\begin{figure}[!h]
\begin{center}
\includegraphics[width=8.5cm] {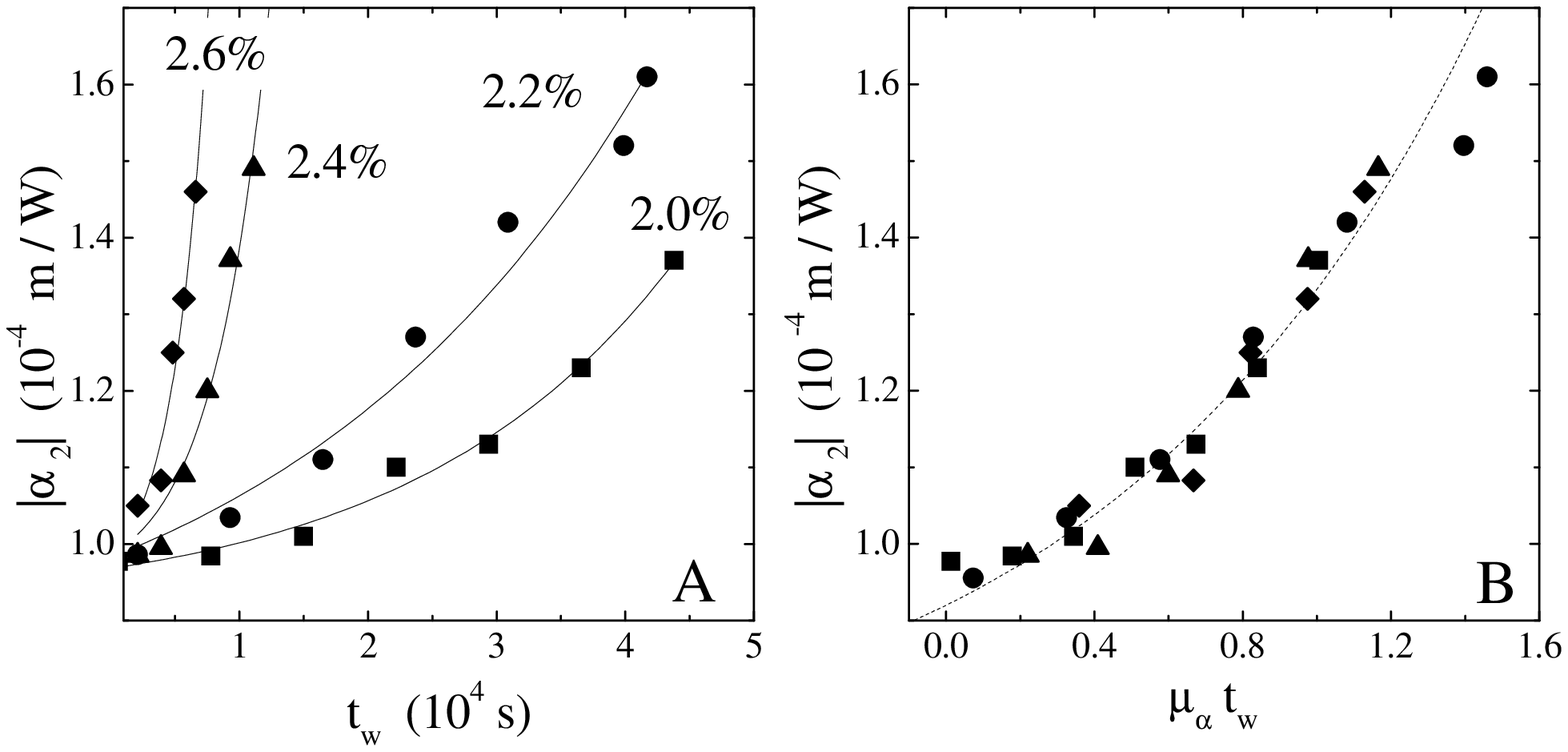}
\caption{Nonlinear absorption coefficient $\alpha_2$ vs. waiting
time $t_w$ (A) and vs. the scaled variable $\mu_\alpha\, t_w$ (B)
at various clay concentrations; continous lines in (A) are the
fitted curves.} \label{fig4}
\end{center}
\end{figure}

The transmittance curves shape in Fig.~\ref{fig1}B corresponds to
a defocusing nonlinear system ($n_2 < 0$) as due to thermal
lens-like behavior, well known in dye solutions. Other effects, as
for example the electrostrictive one, can be ruled out as it would
produce a focusing behaviour ($n_2 > 0$). Additionally the
expected electrostricitve contribution to $n_2$ for laponite
platelets is expected to be at least two or three order of
magnitudes less than those competing to spherical particles
\cite{Ash82a}, as due to the smaller particle volume, i.e. many
order of magnitudes of the defocusing $n_2$ measured in our
samples ($|n_2| \cong 10^{-10}$ m$^2$ W$^{-1}$).  To fit the
experimental data we used the analytical description of Z-scan
method reported in \cite{Sam98b}, based on the Fresnel-Huygens
principle, here enriched by including sample nonlinear absorbance
effects on transmittance, not considered since now.
Previous investigations concerning thermal lensing effects, as
e.g. in Ref.~\cite{Wu90JAP} have shown that the radial temperature
profile adiabatically follows the beam intensity close to the beam
centre, while it slowly decays in the tails, due to thermal
diffusion.  In this case the problem can be treated
perturbatively, by first determining the temperature profile from
the input beam, and then determining the beam diffraction out of
the sample, as it is the case here. As a result the following
equations govern laser beam phase $\Phi(\zeta)$ and intensity
$I(\zeta)$ variations inside a nonlinear sample:
\begin{equation}\label{dI}
\frac{\mbox{d} I(\zeta)}{\mbox{d}\zeta}=-\alpha(I_0)\,I(\zeta)
\end{equation}
\begin{equation}\label{ddeltafi}
\frac{\mbox{d} \Phi (\zeta)}{\mbox{d}\zeta}=k \,
n(I(\zeta))\text{,}
\end{equation}
where $\alpha(I_0)=\alpha_0+\alpha_2 I_0$, with $\alpha_0$, the
linear absorption coefficient, and $\alpha_2$ the nonlinear
coefficient, $n(I(\zeta))=n_0+n_2 I(\zeta)$, with $n_0$, the
linear refractive index, and $n_2$ the Kerr coefficient. In
(\ref{dI}) and (\ref{ddeltafi}) $\zeta$ is the coordinate inside
the sample along the $z$-direction and $k$ the wavenumber. Using
the expression for the transmittance reported in \cite{Sam98b}, by
solving equations (\ref{dI}) and (\ref{ddeltafi}) for  a
$TEM_{00}$ electric incident field intensity distribution and by
determining  the electric {\it far field} $E_C$, transmittance results as:
\begin{equation}\label{T}
T=\frac{|E_C(z,\Phi_o(z))|^2}{|E_C(z, 0)|^2}= \left|\frac{\xi(z)
\Gamma \left( \xi(z),\, i \Phi_o(z)\right)}{\left[i \Phi_o
(z)\right]^{\xi(z)}}\right|^2\text{.}
\end{equation}
In Eq. (\ref{T}) $\Gamma(x,y)$ is the lower incomplete gamma
function,
\begin{equation}\label{xi}
\xi(z)=\frac{1}{2}\left[\frac{i}{z_0}\left(z+\frac{z^2+z_0^2}{D-z}\right)+1\right],
\end{equation}
\begin{equation}\label{deltafi}
\Phi_o(z)=\frac{k \,n_2\, \Lambda\,I_o}{1+z^2/z_0^2}
\end{equation}
being $z_o=2\,k\,w_o^2$, $D$ the distance between the observation
point and the beam waist and $\Lambda$ the effective sample
thickness $\Lambda=[1-exp\,(-\alpha(I_o)\,L)]/\alpha(I_o)$, with
$L$ the actual sample thickness.
We used (\ref{T}) as fitting expression to estimate both $n_2$ and
$\alpha_2$; the latter is shown in Fig.~\ref{fig4}A vs. $t_w$ for
the different clay concentrations, while from our analysis $n_2$
values (not reported) are practically unchanged during the aging
time. Indeed the nonlinear refractive index coefficient $n_2$ is
less sensitive to matter structural changes because it mainly
originates from the water density profile arising from the
temperature profile \cite{Sin00JAP}, a quantity that only slightly
changes during aging. This is not the case for $\alpha_2$, as the
water absorption coefficient is negligible and the measured
$\alpha$ originates only from the Rh-B/Laponite complex. On the
other hand, from Fig.~\ref{fig4}A it is evident that the nonlinear
absorption ($\alpha_2$) is clearly influenced by clay structural
evolution.
According to the {\it theoretical analysis} reported below, we
fitted the Z-scan curves using the function
$\alpha_2=a+b\,\mbox{exp}\,(\mu_\alpha\, t_w)$, with $a$, $b$ and
$\mu_\alpha$ as free parameters. The fitting curves are reported
in Fig.~\ref{fig4}A as full lines. In Fig.~\ref{fig4}B the same
data collapsed on a single master curve are reported as a function
of the scaled time $t_w \mu_\alpha$. The time scaling parameters,
$\mu_\tau$ and
$\mu_\alpha$, obtained respectively from DLS and Z-scan data, are
shown in Table~\ref{tab1}.
\begin{table}[h]
\caption{Time scaling parameters} \label{tab1}
\begin{center}
\begin{tabular}{ccc}
\hline \hline
Clay $wt\%$  &  $\mu_\tau$                  &  $\mu_\alpha$             \\
\hline
$2.0\,$        &   $(1.9\pm 0.2)\,10^{-5}$  &  $(2.3\pm 0.2)\,10^{-5}$  \\
$2.2\,$      &   $(3.5\pm 0.1)\,10^{-5}$    &  $(3.3\pm 0.2)\,10^{-5}$  \\
$2.4\,$      &   $(10.5\pm 0.2)\,10^{-5}$   &  $(10.9\pm 0.5)\,10^{-5}$ \\
$2.6\,$      &   $(28\pm 1)\,10^{-5}$       &  $(17\pm 1)\,10^{-5}$     \\
\hline \hline
\end{tabular}
\end{center}
\end{table}
It can be noticed that $\mu_\tau$ and $\mu_\alpha$ values are
comparable except for the highest concentration, as most likely
due to the different durations of each experiment (i.e. $5$
minutes for photocorrelation and about $15$ minutes for Z-scan
single measurement), which are not negligible in the presence of a
fast gelation process.

{\it Theoretical analysis.} At the microscopic level, the observed
data can be interpreted by taking into account the so called
Ludwig-Soret effect, \cite{Rus04aJOSA} i.~e. the Laponite
platelets concentration ($c$) gradient induced by the presence of
a temperature gradient. Specifically, as the value of $\nabla c$
depends -at fixed $\nabla T$- \cite{note} on mass self diffusion
coefficient, in turn strongly related to the aging time, we expect
a waiting time dependence of $\nabla c$. This mass diffusion
process strongly affects the Rhodamine molecule concentration due
to their adsorption \cite{Arb98aLang} on clay disks. Indeed, we
estimated about $60$ Rh-B molecules for each Laponite platelet,
much less that the maximum allowed adsorption sites (about $500$
sites as provided by Laporte Ltd. datasheet); consequently all dye
particles in solution can be considered as attached to the
colloids.

The previous argument can be put on a quantitative basis by
considering the absorption as due to $N(t_w,I_o)$ dye particles
varying in time due to matter and thermal diffusion processes and
yielding an overall absorption coefficient
\begin{equation}\label{Nalfa1}
\bar{\alpha}(t_w,I_o)=N(t_w,I_o)\,\alpha(I_o).
\end{equation}
Being $\alpha=\alpha_o+\alpha_2\,I_o$, expanding $N(t_w,I_o)$ at
first order in $I_o$, and retaining only the terms linear in
$I_o$, equation (\ref{Nalfa1}) reads as
\begin{equation}\label{Nalfa2}
\bar{\alpha}(t_w,I_o)=N_o\,\alpha_o+ \left[ N_o \alpha_2 +
\alpha_o \, \left( \frac{\partial  N}{\partial\,I_o} \right)
\right] \, I_o,
\end{equation}
From Eq.~(\ref{Nalfa2}), the time dependent nonlinear absorption
coefficient is then written as:
\begin{equation}\label{alfa2bar}
\bar{\alpha}_2(t_w)=N_0\,\alpha_2+\alpha_0(\partial N /\partial
I_0)
\end{equation}
which undergoes aging because of the $t_w$ dependence of
$\left({\partial N}/{\partial I_o} \right)$, i.~e. because of the
variation of particles number due to diffusive effects. To
determine the waiting time dependence of $\partial N /\partial
I_o$ in (\ref{alfa2bar}), we rely on the stationary solutions of
both heat equation and mass flow continuity equation reported in
\cite{Rus04aJOSA}, which give the following radial profile for the
concentration $c(r,t_w)$ ($\propto N(r,t_w)$):
\begin{equation}\label{Deltac}
\nabla^2 c(r,t_w)= \frac{2\,L\,I_0}{\kappa}
\frac{D_T}{D(t_w)}\,\bar{c}\,(1-\bar{c})\,e^{-2\,r^2/w_0^2}
\end{equation}
being $\kappa$ the thermal conductivity,
$2\,L\,I_0\,\mbox{exp}(-2\,r^2/w_0^2)$ the Gaussian beam source
term, $\bar{c}$ the average colloids concentration, $D(t_w)$ the
mass self diffusion coefficient and $D_T$ the thermal diffusion
coefficient. In writing equation (\ref{Deltac}) we assume that the
concentration gradient time dependence is enclosed uniquely in
mass diffusion coefficient, while $D_T$ was assume to be
unchanged, and all other terms are constant. Solution of
Eq.~(\ref{Deltac}) yields, besides an integration constant, the
following radial concentration profile
\begin{eqnarray}\label{Deltac1}
c(r,t_w) \!=\! -\frac{D_T L I_o w_0^2 \bar{c} (1-\bar{c})}{4
\kappa D(t_w)} \left[ E_i(\frac{-2 r^2}{w^2})-2\,log(r) \right ]
\end{eqnarray}
where $E_i(x)$ denotes the exponential integral function. The
maximum difference in concentration within the sample  in a
confined geometry of thickness $L$ given by (\ref{Deltac1}) is:
\begin{equation}\label{Deltac2}
\Delta
c(t_w)=\frac{D_T\,L\,I_0\,w_0^2\,\bar{c}\,(1-\bar{c})}{4\,\kappa\,D(t_w)}\,[\gamma+2\,log(H)],
\end{equation}
with $\gamma$ the Euler constant; in writing (\ref{Deltac2}) we
regard the logaritmic and the exponential terms as main
contributions to $\Delta c(t_w)$ for large ($r\simeq H$, being $H$
the transverse sample cell dimension) and for small ($r\simeq
0$) distances respectively.

Finally, from Eqs.~(\ref{Deltac2}) and (\ref{alfa2bar}),
\begin{equation}
\label{alfa2barbis}
\bar{\alpha}_2(t_w)=N_0\,\alpha_2-C\,\frac{\alpha_0}{D(t_w)}
\end{equation}
where the constant $C$ encloses all time independent terms in
(\ref{Deltac2}) and it also takes account of the proportionality
relation $c(t_w)\propto N(t_w)$. Expression (\ref{alfa2barbis})
gives the $t_w$ dependence of  $\bar{\alpha}_2$, and, noticing
that $\mu_\tau \approx \mu_\alpha$, one can conclude that $D
\propto \tau_m^{-1}$. This proportionality suggests that the
gelation process proceeds through the structuring of
interconnecting clay platelets network rather than through
clusters growth and aggregation.

In conclusion, we studied the aging of the non-linear optical
susceptibility in a water-RhB-clay solution. The dye molecules are
adsorbed on the clay surface, creating a complex medium with a
high optical nonlinearity capable to perform a liquid-gel
transition on a time scale $\mu_\tau^{-1}$, dependent on the clay
concentration. To analyze the waiting time dependence of the
Z-scan curves, we derived Eqs.~(\ref{T}-\ref{deltafi}), which
generalize the result obtained in Ref.~\cite{Sam98b} to the case
where photon absorption is present. At variance with the linear
optical properties, which are not affected by aging, we showed
that {\it i)} the non-linear optical susceptibility actually ages,
{\it ii)} its $t_w$ dependence is only function of a concentration
dependent time scale $\mu_\alpha^{-1}$, {\it iii)} this time scale
is the same as that for the waiting time dependence of the
relaxation time measured by DLS, {\it iv)} the waiting time
dependence can be microscopically explained assuming a
thermodiffusive behavior of the clay platelets (Soret effect), and
{\it v)} the inverse proportionality between $D$ and $\tau_m$
suggests that the structural arrest of the laponite platelets
takes place via the organization and the structuring of a
spatially homogeneous clay network.

We thank A. Martinelli for UV-visible absorption spectra, R. Di
Leonardo for DLS software realization, E. Del Re, R. Piazza,  B.
Ruzicka, F. Sciortino, S. Trillo and L. Zulian for interesting
discussions and useful suggestions.

\begin{thebibliography}{22}
\bibitem{note} It is worth to notice that the temperature increase at the beam
center, in the actual experimental condition, is estimated to be
6-7 K, a negligible value compared to the clay-clay interaction
energy \cite{Dij97PRE}.
\end{thebibliography}

\end{document}